\documentclass[twocolumn,preprintnumbers]{revtex4}
\usepackage{amssymb}
\usepackage{amsmath}
\usepackage{graphicx}
\usepackage{graphics}
\usepackage{dcolumn}
\usepackage{bm}

\begin{document}

\title{Comment on "Polar and antiferromagnetic order in f=1 boson systems"}
\author{C. G. Bao}
\affiliation{Center of Theoretical Nuclear Physics, National Laboratory of Heavy Ion
Accelerator, Lanzhou, 730000, PRC and State Key Laboratory of Optoelectronic
Materials and Technologies School of Physics and Engineering, Sun Yat-Sen
University, Guangzhou, PRC}
\begin{abstract}
An inequality for the lower bound of the average number of
hyperfine component $\mu =0$ particles in the ground state of spin-1
condensates under a magnetic field
has been derived in ref.\cite{tasa13}. It is shown in this comment that, in a broad domain of parameters usually accessed in experiments, the lower bound appears to be negative.
 Thus the applicability of the inequality is very limited.
\end{abstract}
\maketitle

The essential results of Tasaki's letter \cite{tasa13} are concluded
in two theorems. In particular, an inequality (eq.(11) of
\cite{tasa13} ) is derived to constrain the number of spin-component
$\mu =0$ particles in the ground state (g.s.). The inequality has
the following short-comings.

(i) Eq.(11)of \cite{tasa13} can be rewritten as
\begin{equation}
\langle \Phi _{GS},\hat{\rho} _{0}\Phi _{GS}\rangle \geq 1-c_{0}\frac{N}{2qV}%
,
\end{equation}
where $c_{0}=(2g_{2}+g_{0})/3$ is the strength of the central
(spin-independent) force. Thus, this inequality demonstrates that the
constraint arises essentially from the central force. This is misleading.
Since the spin-flip is caused by the spin-dependent force with the strength $%
c_{2}=(g_{2}-g_{0})/3$, it should be $c_{2}$ to play the essential role.

(ii) There is a great difference between the ground states of Rb
($c_{2}<0$) and Na ($c_{2}>0$) condensates (Say, when $q=0$, the
spins of Rb are coupled to total spin $S=N$, while the spins of Na
are coupled to $S=M$, where $M$ is the total magnetization). It has
been shown by numerical calculation that the variations of the
$\langle \Phi _{GS},\hat{\rho}_{0}\Phi _{GS}\rangle $ versus $q$ of
these two species differ from each other greatly (refer to Fig.2b
and Fig.3b of \cite{bao12}, where the scales for $q$ are greatly
different). Therefore, the constraints for the two species should be
greatly different.

(iii) When a magnetic field is applied, the g.s. is not necessary to have
 $M=0$ (In fact, the energy of the states with a larger $M$ may be remarkably reduced
 due to the negative linear Zeeman energy). It has been shown that $\langle \Phi _{GS},\hat{\rho} _{0}\Phi
_{GS}\rangle $ in fact depends on $M$ seriously.\cite{bao12} The
inequality in \cite{tasa13} is confined only for the case of $M=0$.
A better constraint should take the $M-dependence$ into account.

(iv) The constraint given by eq.(11) of \cite{tasa13} is too loose.
To evaluate numerically, it is assumed that the Na atoms are trapped
by a harmonic potential $%
U(r)=\frac{1}{2} m\omega ^{2}r^{2}$. Then the g.s. of the
single-particle hamiltonian is just the lowest harmonic oscillator
state. When $\hbar \omega $ and $\sqrt{\hbar /m\omega }$ are used as
units for energy and length, according to the definition given in
\cite{tasa13}, $1/V=(1/\pi )^{3/2}$. When $q$ and $c_{0}$ are
expressed in the new units, their values are
$q=1745(B/G)^{2}/(\omega \sec )$ and $c_{0}=6.77\times
10^{-4}(\omega \sec )^{1/2}$, where $G$ is Gauss and $B$ is the
magnetic field ($B/G$ and $\omega \sec $ are both a number).
Accordingly, eq.(11) of \cite{tasa13} becomes
\begin{equation}
\langle \Phi _{GS},\hat{\rho}_{0}\Phi _{GS}\rangle \geq 1-3.48\times 10^{-8}%
\frac{N(\omega \sec )^{3/2}}{(B/G)^{2}}
\end{equation}%
Since $N$ is usually large, the right side might become negative.
For an
example, if $N=10^{4}$ and $\omega =300\sec ^{-1}$, the right side becomes $%
1-\frac{1.81}{(B/G)^{2}}$. It implies that, when $B<1.34G$, the
right side is negative. Thus, in a large domain quite often
accessed, the inequality makes no sense.

In the derivation of Theorem 2 the contribution from the term
$\langle \Phi _{GS}, \hat{V} \Phi _{GS}\rangle $ has been completely
neglected. In general, during the derivation of a formula or an
inequality, the neglect of a small term is allowed. However, At the
present case, $\langle \Phi _{GS}, \hat{V} \Phi _{GS}\rangle $ is
not a small term. Instead, it is essential. In order to overcome the
above shortcomings, this term should be recovered.

Incidentally, the statement stated in \textit{Theorem 1} is not new. A more
general formula for $\langle \vartheta _{SM}^{[N]},\hat{\rho}_{\mu
}\vartheta _{SM}^{[N]}\rangle $ has been given in Eq.(10) of Ref.\cite{bao12}%
. When $\mu =0$, the general formula can be rewritten as
\begin{equation}
\langle \vartheta _{SM}^{[N]},\hat{\rho}_{0}\vartheta _{SM}^{[N]}\rangle =%
\frac{S(S+1)(2N+1)-N-M^{2}(2N+3)}{N(2S+3)(2S-1)}.
\end{equation}

When $S=M=0$, the above formula gives $\langle \vartheta _{00}^{[N]},\hat{%
\rho}_{0}\vartheta _{00}^{[N]}\rangle =1/3$ as stated in \textit{Theorem 1}.

\end{document}